\documentclass[prc,twocolumn,superscriptaddress,showpacs,floatfix,tightenlines]{revtex4}
\usepackage{graphicx,amsmath,amssymb,bm}
\usepackage{amsfonts}
\usepackage{hyperref}
\usepackage[labelfont={normalsize},subrefformat=parens,caption=false]{subfig}
\newcommand{\beq}{\begin{equation}}
\newcommand{\eeq}{\end{equation}}
\newcommand{\bea}{\begin{eqnarray}}
\newcommand{\eea}{\end{eqnarray}}

\newcommand{\ts}{\textstyle}

\newcommand{\fmi}{\, \text{fm}^{-1}}

\newcommand{\Nmax}{\ensuremath{N_{\rm max}}}
\newcommand{\hw}{\ensuremath{\hbar\Omega}}

\newcommand{\kinf}{\ensuremath{k_{\infty}}}

\newcommand{\Einf}{\ensuremath{E_{\infty}}}
\newcommand{\Ainf}{\ensuremath{A_{\infty}}}

\newcommand{\rsqav}{\langle r^2 \rangle}
\newcommand{\rsqinf}{\ensuremath{\rsqav_{\infty}}}

\newcommand{\LamUV}{\Lambda_{UV}}

\begin{document}

\title{Corrections to nuclear energies and radii in finite oscillator spaces}

\author{R.~J.~Furnstahl}
\affiliation{Department of Physics, The Ohio State University, Columbus, Ohio 43210, USA}
\author{G.~Hagen}
\affiliation{Physics Division, Oak Ridge National Laboratory,
Oak Ridge, Tennessee 37831, USA}
\affiliation{Department of Physics and Astronomy, University of
Tennessee, Knoxville, Tennessee 37996, USA}
\author{T.~Papenbrock}
\affiliation{Department of Physics and Astronomy, University of
Tennessee, Knoxville, Tennessee 37996, USA}
\affiliation{Physics Division, Oak Ridge National Laboratory,
Oak Ridge, Tennessee 37831, USA}

\begin{abstract}
  We derive corrections to the ground-state energies and radii of
  atomic nuclei that result from the limitations of finite oscillator
  spaces.
\end{abstract}


\maketitle

{\it Introduction} -- The oscillator basis is widely used in nuclear
structure computations because it allows the practitioner to exploit
and implement all symmetries of the nuclear many-body problem, and
because its localized nature corresponds well to the structure of the
self-bound atomic nucleus. After all, the nuclear shell model is based
on the harmonic oscillator with a strong spin-orbit
splitting~\cite{MJ}. Several computational implementations of {\it ab
  initio} methods~\cite{zheng1993,navratil2009} and nuclear density
functional theory~\cite{dobaczewski1997} essentially start from the
oscillator basis. Basic observables sought in such computations
include the binding energies and radii.  Ideally, the computed
observables should be independent of the parameters of the employed
oscillator space, i.e., the maximum number of oscillator quanta $N$ and
the frequency $\Omega$ of the oscillator wave functions.  This ideal
is often difficult to reach in practice, so various empirical
extrapolation
schemes~\cite{hagen2007,bogner2008,forssen2008,maris2009,roth2009}
have been applied, but all lack a firm theoretical foundation.

The proper accounting for corrections to nuclear energies and radii
that arise in finite oscillator spaces is an important problem for
several reasons. First, a theoretical foundation of such corrections
would enable the practitioner to extrapolate reliably from smaller
model spaces and thus extend the reach of some computational methods.
This is particularly important for weakly bound nuclei where the
Gaussian falloff of the oscillator basis can capture a halo state
often only in unachievable large model spaces~\cite{bogner2008}.
Second, uncertainty quantification of results -- standard in
experimental research -- is increasingly taking place in nuclear
structure theory~\cite{bogner2010}.  Here, the quantification of
theoretical uncertainties due to the nuclear interaction is possible
for interactions from effective field theory (EFT), but the robust
quantification of errors due to finite oscillator spaces is lacking.
Finally, important steps towards an harmonic-oscillator based EFT for
the nuclear shell model have been made
recently~\cite{haxton2000,stetcu2007,coon2012}.  Such a theory should
also control and exploit the limitations of the finite model space.

In this Rapid Communication, we derive corrections of nuclear energies
and radii that are due to finite oscillator spaces.  We build on the
insights of Coon {\it et al.}~\cite{coon2012}, who focus on the
infrared and ultraviolet cutoffs induced by a truncated basis.  Our
derivations are based on simple arguments and verified in a
one-dimensional model.  We apply the results to $^{16}$O and
demonstrate that the theoretical corrections agree well with the
numerical data.  Calculations for the $^6$He ground-state energy and
neutron radius show that predictions are feasible even for halo nuclei.

{\it Theoretical derivation} -- For a particle in a box with
periodic boundary conditions, L{\"u}scher
derived the corrections to bound states due to the finite size of the
box~\cite{luscher1986}.  Our derivation is analogous, 
except that the size of the box is now given in terms of the
spatial extension of the oscillator basis and we deal
essentially with Dirichlet boundary conditions.  Let us consider a
model space of oscillator wave functions with maximum oscillator
energy $E=\hw(N+3/2)$. In practice, one has to choose $\hw$ and
$N$ such that the momentum cutoff $\lambda$ of the employed
interaction is smaller than the ultraviolet (UV) momentum
\beq 
\label{lamUV}
\Lambda_{UV}\equiv
\sqrt{2(N+3/2)}\hbar/b \ , 
\eeq 
and that the radius $r$ of the nucleus is smaller than the radial
extent
\beq 
\label{L}
L_0\equiv \sqrt{2(N+3/2)} b
\eeq 
of the employed oscillator space. Here, $b\equiv\sqrt{\hbar/(m\Omega)}$ is
the oscillator length of our basis, and $m$ denotes the nucleon mass.
The definitions (\ref{lamUV}) and (\ref{L}) are indeed the maximum
momentum and displacement, respectively, of a particle in a harmonic
oscillator at energy $E=\hw(N+3/2)$. They differ from previous scaling 
relations~\cite{hagen2010a,jurgenson2011} by factors of $\sqrt{2}$.

In practice, satisfying the UV condition $\lambda < \Lambda_{UV}$ and
the infrared (IR) condition $r<L_0$ does not guarantee converged
nuclear structure results in the oscillator basis because the momentum
cutoff $\lambda$ is usually not sharp, and the nuclear wave function
extends beyond the nuclear radius $r$. However, nuclear interactions
from chiral EFT and from renormalization group transformations exhibit
a super-Gaussian falloff in momentum space, whereas the nuclear wave
function only falls off exponentially in coordinate space. Thus, once
$\lambda < \Lambda_{UV}$ holds, the UV convergence in momentum space
will be rapid, and one is dominated by corrections from the slower
falloff in coordinate space.  Practitioners of nuclear structure
computations know this very well (see, e.g., Fig.~\ref{fig6} below):
When energies are plotted as a function of $\hw$, the minimum
initially shifts toward larger values of $\hw$ as $N$ is
increased. However, once UV convergence has been reached, further
increasing $N$ shifts the minimum back to lower values of $\hw$ to
capture the coordinate-space tail of the wave function. In what
follows, we will assume UV convergence and compute the correction from
incomplete IR convergence.

The finite extent of the oscillator basis up to a radius $L$ in
coordinate space essentially requires the wave function to vanish at
$r\approx L$. The maximum radius $L_0$ from Eq.~\eqref{L} is only a leading-order 
(or asymptotically valid) estimate because the oscillator wave function
decays rapidly beyond the classical turning point.
An improved estimate for $L$
using the intercept of the tangent at $r=L_0$
is~\cite{inprep}
\beq
   L \approx L_0 + 0.54437\, b\, (L_0/b)^{-1/3}
   \;,
\eeq
which we use with the analogous expression
for $\LamUV$ in numerical examples below.
For our derivation of IR corrections, we adapt the
discussion in Ref.~\cite{djajaputra2000}.
Given a boundary condition at $r = L$ beyond the range of the nuclear
potential, we write the energy compared to that for $L = \infty$ as
\beq
\label{eq:erg}
  E_L = \Einf + \Delta E_L
  \;,
\eeq
and we seek an estimate for
$\Delta E_L$, which is assumed to be small.

Let $u_E(r)$ be the radial solution with regular
boundary condition at the origin and energy $E$.  We denote the
particular solutions $u_{E_L}(r) \equiv u_L(r)$ and
$u_{\Einf}(r) \equiv u_\infty(r)$.  Then the linear
energy approximation is (for $r \leq L$)~\cite{djajaputra2000}
\beq
  u_L(r) \approx u_\infty(r) 
     + \Delta E_L \left.\frac{du_E(r)}{dE}\right|_{\Einf}
     \;,
     \label{eq:uL}
\eeq
assuming a smooth expansion of $u_E$ about $E = \Einf$ at fixed $r$.
Evaluating Eq.~\eqref{eq:uL} at $r=L$ with the boundary condition
$u_L(L) = 0$, we find
\beq
\Delta E_L \approx
   -u_\infty(L)
   \left(\left.\frac{\ts du_E(L)}{\ts dE}\right|_{\Einf}\right)^{-1}
  \;,
  \label{eq:DeltaEL-estimate}
\eeq
which is the estimate we seek.  For general $E$, the asymptotic form
of the radial wave function for $r$ greater than the range $R$ of the
potential is
\beq
  u_E(r) \stackrel{r \gg R}{\longrightarrow}
    A_E (e^{-k_E r} + \alpha_E e^{+k_E r})
    \;,
    \label{eq:uE}
\eeq
with the known case $u_\infty(r) \stackrel{r \gg R}{\longrightarrow}
\Ainf e^{-\kinf r}$ for $E = \Einf$. Here, $\kinf$ is determined by the nucleon 
separation energy 
\beq
\label{S}
S=\frac{\hbar^2 k_\infty^2}{2m} \ .
\eeq 
We take the derivative of Eq.~\eqref{eq:uE} with
respect to energy, evaluate at $E=\Einf$ using $\alpha_{\Einf} = 0$ and
$dk_E/dE = -m/(\hbar^2 k_E)$ to find
\beq
  \left.\frac{du_E(r)}{dE}\right|_{\Einf} =
  + \Ainf \left.\frac{d\alpha_E}{dE}\right|_{\Einf} e^{+\kinf r}
+ \mathcal{O}\left(e^{-\kinf r}\right)
  \;.
  \label{eq:dalphaE-dr}
\eeq
Substituting Eq.~\eqref{eq:dalphaE-dr} at $r=L$ 
into Eq.~\eqref{eq:DeltaEL-estimate}, we obtain
\bea
\Delta E_L 
   &\approx& -\left[\left.\frac{\ts d\alpha_E}{\ts dE}\right|_{\Einf}\right]^{-1}
   e^{-2\kinf L}   + \mathcal{O}(e^{-4\kinf L})
   \;.
   \label{eq:DeltaEL-final}
\eea  
The prefactor in the square brackets depends on details of the
interaction (but not on $L$), and will be fit to numerical data when
Eq.~\eqref{eq:DeltaEL-final} is used together with Eq.~\eqref{eq:erg}. 
Thus, the main result is 
\beq
\label{E}
E_L = \Einf + a_0 e^{-2\kinf L} \ , 
\eeq
and in practical applications one can treat $\Einf$, $a_0$ and $\kinf$
(in cases where the separation energy is not known) as fit parameters.
Note that our result~(\ref{E}) explains the exponential decay observed
empirically in Ref.~\cite{coon2012}.
In contrast to the L\"uscher result in which the energy is always lowered
by periodic images of the potential~\cite{luscher1986}, the energy
from Eq.~\eqref{E} is always increased by the shift of a node from
$r=\infty$ to $r=L$, consistent with the variational nature of the
truncated basis expansion.

Let us next turn to radii. It is convenient to express the radius
squared as the infinite-model-space result plus a correction term
\beq
\label{eq:r2}
  \rsqav_L = \rsqinf + \Delta\rsqav_L
  \;,
\eeq
where
\beq
  \Delta\rsqav_L = 
  \frac{\int_0^L |u_L(r)|^2\, r^2\,dr}{\int_0^L |u_L(r)|^2\, dr}
   - \frac{\int_0^\infty |u_\infty(r)|^2\, r^2\,dr}{\int_0^\infty |u_\infty(r)|^2\, dr}
   \;.
   \label{eq:Delta-rsq}
\eeq   
Because the dependence on $L$ of $u_L(r)$ in Eq.(5) is confined to
$\Delta E_L$, when $u_L$ is substituted into Eq.~\eqref{eq:Delta-rsq}
the $L$ dependence of each separate integrand comes entirely from the
upper integration limit.  Therefore we can use the asymptotic
expressions $u_{\infty}(r) \longrightarrow \Ainf e^{-\kinf r}$ and
\beq
  \left.\frac{du_E(r)}{dE}\right|_{\Einf} 
  \approx - \frac{\Ainf}{\Delta E_L} e^{-2\kinf L} e^{+\kinf r}  
\eeq
to identify the leading-order expression 
$\Delta\rsqav_L  \propto \rsqinf(2\kinf L)^3 e^{-2\kinf L}$. 
(Note that any $L$-independent terms are guaranteed to cancel
by the definition Eq.~\eqref{eq:Delta-rsq}.)
The next-to-leading-order expression scales as
$(2\kinf L)\exp{(-2\kinf L)}$ because the condition $u_L(L)=0$ ensures
there is no quadratic term in $2\kinf L$.  
Thus, the $L$ dependence of the squared radius is 
(with $\beta\equiv 2\kinf L$)~\cite{inprep}
\bea
 \rsqav_L \approx {\rsqinf}[ 1 - (c_0 \beta^3 +c_1 \beta) e^{-\beta}]
   \;.
\label{rad}
\eea
Here, $\rsqinf$, $c_0$, and $c_1$ are fit parameters while $\kinf$
should be determined in fitting the energy~(\ref{E}). 
The approximation~(\ref{rad}) is valid
in the asymptotic regime $\beta\gg 1$. In practice, one needs
$\beta\gtrsim 3$ because the leading-order correction has its maximum at
$\beta=3$, and the next-to-leading order corrections is
approximately suppressed by one order of magnitude for $\beta \gtrsim 3$
(with $c_0$ and $c_1$ of order unity).

Equations~(\ref{E}) and (\ref{rad}) are the main
results of this Rapid Communication.  A few comments are in order
before we turn to applications of these results.  Note that we derived
these results in the laboratory system. For the nuclear $A$-body
problem, we could also have exploited the separation of the
center-of-mass coordinate in the oscillator basis and followed a
similar derivation for the $A^{\rm th}$ particle with respect to
the center of mass of the remaining $(A-1)$ particles. This would rescale $L$ and
the momentum of the $A^{\rm th}$ particle accordingly, but the final
results are unchanged when re-expressed in laboratory coordinates.
In light of the approximations involved in defining $L$ and 
the corrections to the energy and radius, 
in actual fits to numerical results one might want to treat $\kinf$ as a
fit parameter even when the separation energy is known.

\begin{figure}[h]
\includegraphics[width=0.48\textwidth,angle=0]{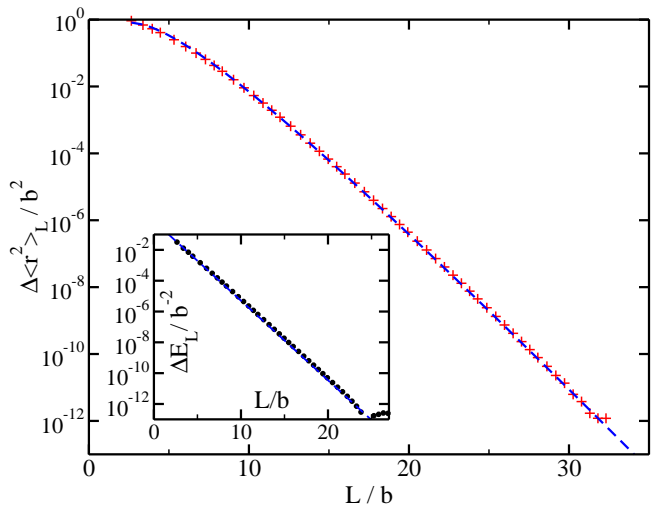}
\caption{(color online) Finite-basis-size correction of the squared radius (crosses) compared
  to Eq.~(\ref{rad}) (dashed line) for a toy model.  The squared radius is
  $r_\infty^2/b^2 \approx 1.736$, and $k_\infty b \approx 0.595$. Inset:
  Finite-basis-size correction of the energy (data points) compared to
  Eq.~(\ref{eq:DeltaEL-final}) (dashed line).}
\label{fig1}
\end{figure}

{\it Applications} -- As a first check, we consider a toy model in one
dimension with the Hamiltonian $H=p^2/2 -v_0\exp{(-x^2)}$. Here, $x$ is
given in units of the oscillator length $b$. In one dimension, the
constant $3/2$ in Eqs.~(\ref{lamUV}) and (\ref{L}) has to be replaced
by $1/2$. We compute the ground-state energy and the squared radius
for $v_0=0.5$ in large oscillator spaces to obtain fully converged
results for the ground-state energy and the radius. In this simple
case, the ground-state energy is given in terms of the separation
energy~(\ref{S}) as $E_\infty=-S$.  Figure~\ref{fig1} shows the
correction $\Delta\langle r^2\rangle$ as a function of $L$. The dashed
line results from a leading-order fit to Eq.~(\ref{rad}), and the agreement between
numerical data and the theoretical prediction extends over ten orders
of magnitude. The inset shows that the $L$-dependent energy correction
also agrees with the prediction~(\ref{E}).

\begin{figure}[h]
\includegraphics[width=0.45\textwidth]{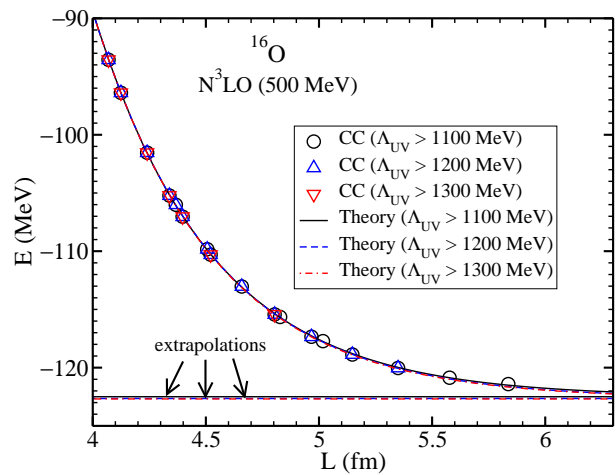}
\caption{(color online) Open symbols: Ground-state energy of $^{16}$O
  as a function of $L$. Lines: Fits to Eq.~\eqref{E} yield
  $E_\infty\approx -122.6$~MeV and $k_\infty\approx 0.95~\fmi$.
  $\Lambda_{UV}$ from Eq.~(\ref{lamUV}).}
\label{fig2}
\end{figure}

Let us turn to the nuclear many-body problem. We employ the
nucleon-nucleon interaction from chiral EFT by Entem and
Machleidt~\cite{entem2003}, and compute the ground-state energy and
radius of the nucleus $^{16}$O with the coupled-cluster method in its
singles and doubles approximation with triples
corrections~\cite{ccm,hagen2010a}. We employ model spaces with
frequencies $42 \le\hw/{\rm MeV} \le 76$ and with $N=12, 14$. To
ensure that the computed results are practically UV converged, we only
use those oscillator spaces for which $\Lambda_{UV}$ is sufficiently
large.  Figure~\ref{fig2} shows the results for the ground-state
energy as a function of $L$. The circles, up triangles, and down
triangles denote points with $\Lambda_{UV}>1100$~MeV,
$\Lambda_{UV}>1200$~MeV, and $\Lambda_{UV}>1300$~MeV, respectively.
The points all fall on a line because UV convergence has practically
been achieved. Thus, we can apply our theory. The lines show fits to
Eq.~\eqref{E} with fit parameters $E_\infty$, $a_0$ and $\kinf$. Note
that the result $E_\infty\approx -122.6$~MeV of the fit depends very
weakly on $\Lambda_{UV}$, the difference being about 0.2~MeV. In the
fits, we obtain $k_\infty\approx 0.95~\fmi$, and this agrees well with
the decay of the $p_{1/2}$ orbital that contributes to the
density~\cite{vanNeck1997}.

\begin{figure}[h]
\includegraphics[width=0.45\textwidth]{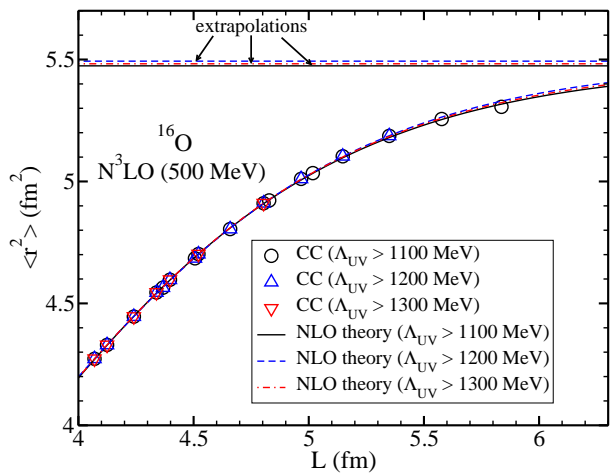}
\caption{(color online) Open symbols: Squared radius as a function of $L$ for
  $^{16}$O. Lines: Fits of Eq.~(\ref{rad}) with $k_\infty$
  fixed from the energy fit.  $\Lambda_{UV}$ from
  Eq.~(\ref{lamUV}).}
\label{fig3}
\end{figure}

Next we consider the radius. We use Eq.~(\ref{rad}) including the
next-to-leading order correction and fit the parameters $\rsqinf$,
$c_0$, and $c_1$ to data, with $\kinf$ taken from the fit of the
ground-state energy. The result is shown in Fig.~\ref{fig3}. The
circles, up triangles, and down triangles denote points with
$\Lambda_{UV}>1100$~MeV, $\Lambda_{UV}>1200$~MeV, and
$\Lambda_{UV}>1300$~MeV, respectively. The lines show the
corresponding fits and asymptotes, and the extrapolated radius is
$r\approx 2.34$~fm. It is particularly
satisfying that the extrapolation also works well for the few data
points with $\Lambda_{UV}>1300$~MeV.

We also consider the challenging case of a halo nucleus. The
isotope $^6$He is only bound by about $0.97$~MeV with respect to
$^4$He, and thus exhibits a two-neutron halo. Note that $^5$He is not
a bound nucleus, and that the neutron separation energy of $^6$He is
about $1.86$~MeV. As a consequence of the weak binding, the matter
radius of $^6$He is unusually large (about 2.4~fm compared to 1.5~fm
for $^4$He)~\cite{tanihata1992,alkhazov1997,kiselev2005}. 
We address this challenge by applying the
finite-basis-size corrections to the energy~(\ref{E})
and neutron radius~(\ref{rad}). 

\begin{figure}[h]
\includegraphics[width=0.4\textwidth]{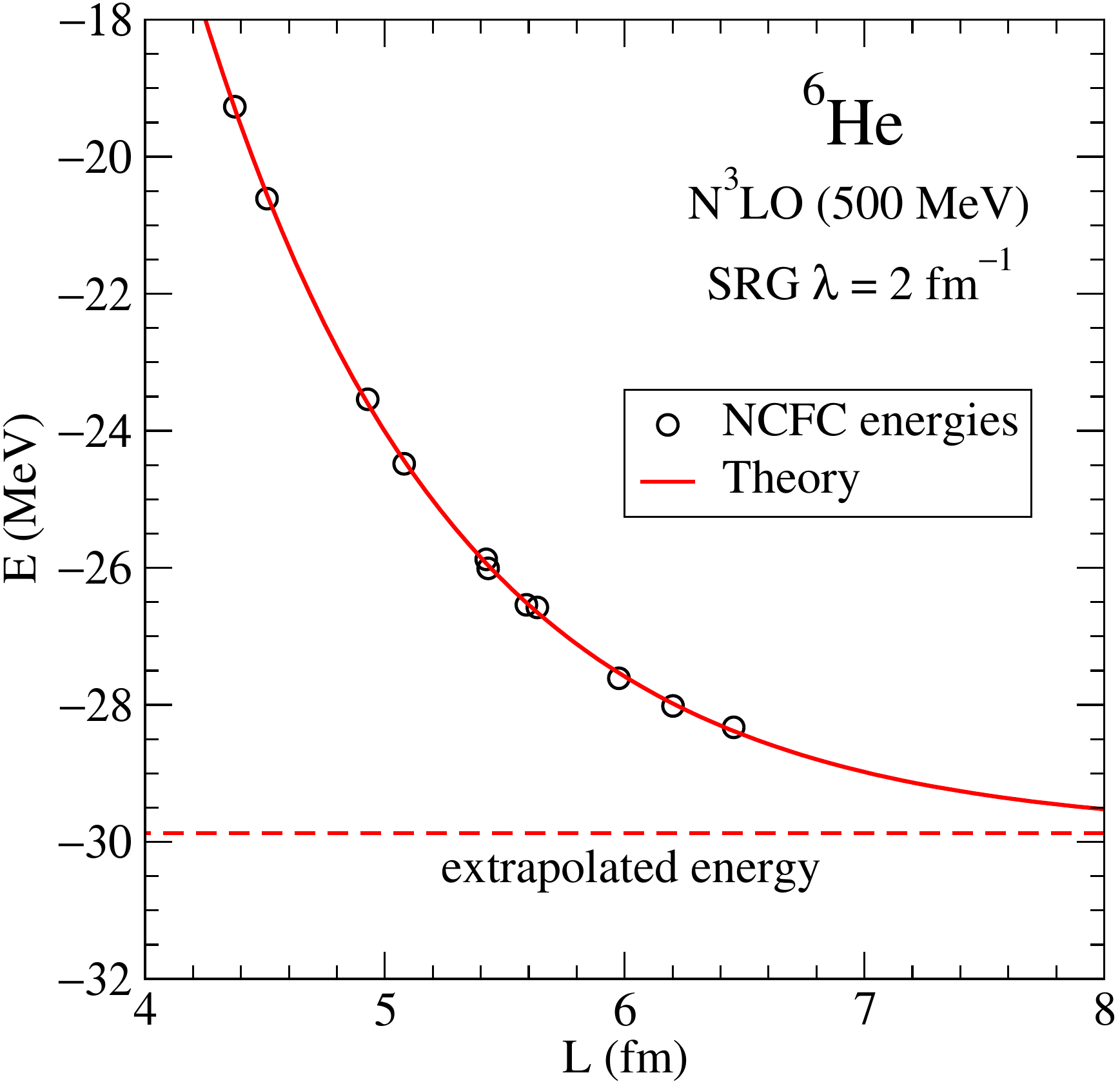}
\caption{(color online) Circles: NCFC ground-state energies of the
  halo nucleus $^6$He from Ref.~\cite{bogner2008}. Full line: Fit of
  Eq.~\eqref{E} yields $\Einf\approx -29.87\,$MeV (dashed line) and
  $\hbar\kinf\approx 93\,$MeV. }
\label{fig4}
\end{figure}

Our test case uses NCFC results~\cite{bogner2008} obtained for a
chiral EFT nucleon-nucleon interaction that was softened via a
similarity renormalization group (SRG)
transformation~\cite{bogner2007} with a parameter $\lambda=2.0\fmi$.
Figure~\ref{fig4} shows the fit of the ground-state energy for model
spaces with $\Lambda_{UV}>660$~MeV and $\hw\geq 24$~MeV (which ensures a
small UV correction). The fit yields $E_\infty\approx -29.87$~MeV, and
the computed two-neutron separation energy is about $0.95$~MeV. Thus,
both energies are in good agreement with experiment (despite the
absence of a three-body force).  The fit also yields $\hbar\kinf\approx
93$~MeV corresponding to a neutron-separation energy of $S\approx
4.6$~MeV, which is significant larger than the experimental value.  Of
course, the unbound nucleus $^5$He cannot be computed reliably in the
oscillator basis, but the interpretation of $\kinf$ in this case
requires further study.

\begin{figure}[h]
\includegraphics[width=0.4\textwidth]{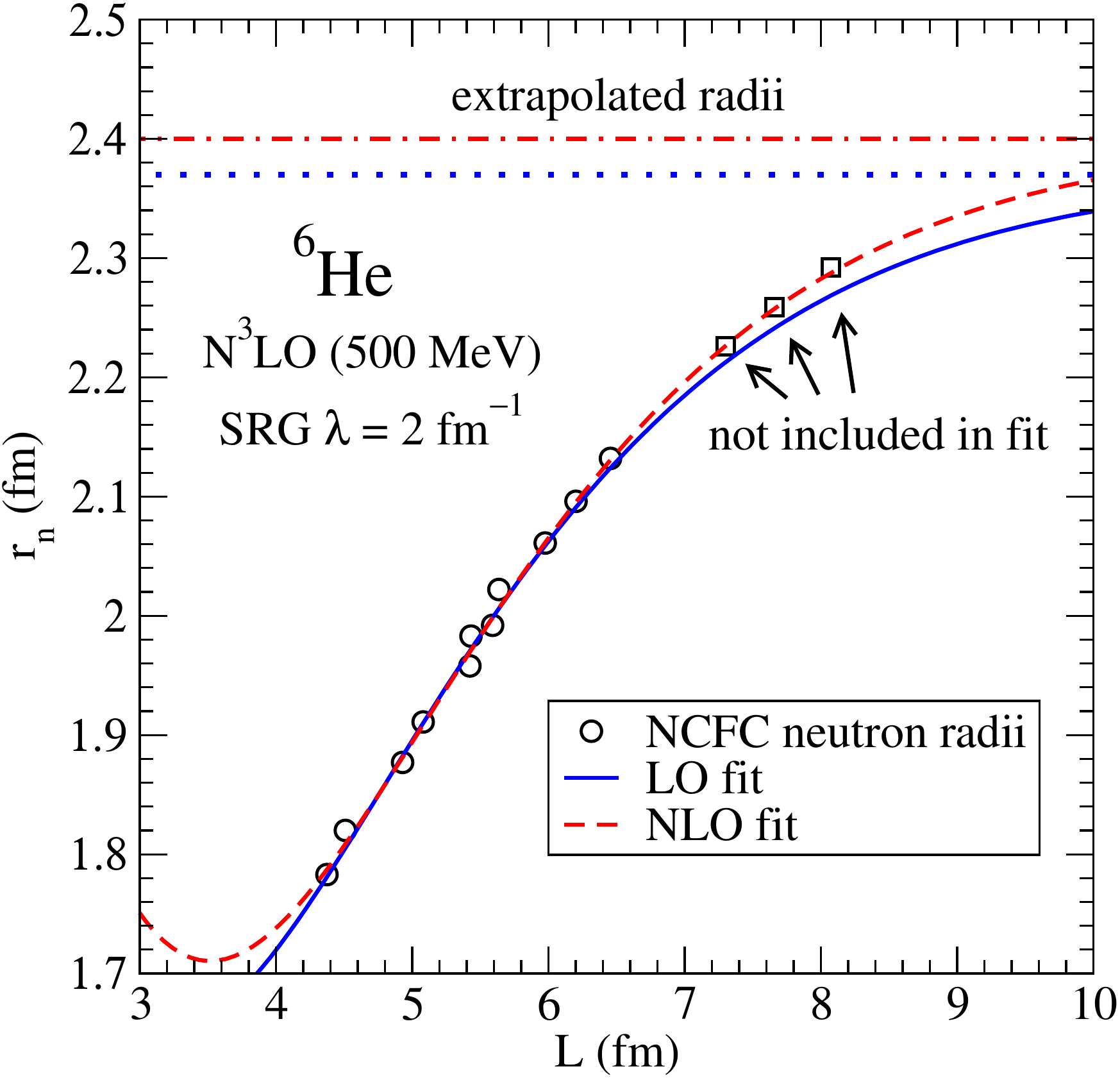}
\caption{(color online) Circles and squares: NCFC neutron radii of
  $^6$He. Lines: Fits of Eq.~\eqref{rad} to circles with $k_\infty$
  fixed from energy fit yield a radius of 2.37\,fm at LO and 2.40\,fm
  at NLO.}
\label{fig5}
\end{figure}

Now we can extrapolate the neutron radius of $^6$He.  The results of
NCFC calculations are shown in Fig.~\ref{fig5}. Without a knowledge of
the finite-basis-size corrections, it would be impossible to make any
reasonable prediction for the radius because of the apparent lack of
convergence.  Note, however, that we are in the UV-converged regime
and $2\kinf L>3$ for the NCFC data points in Fig.~\ref{fig4}.  Thus,
we should be able to apply our correction formula.  The solid and
dashed lines in Fig.~\ref{fig5} show the results for the radius based
on a fit at leading order and next-to-leading order, respectively. At
next-to-leading order we find $r\approx 2.40$~fm, and this prediction
is in reasonable agreement with deductions from
data~\cite{alkhazov1997}.  Several additional points at large $L$ not
included in the fit are in good agreement with the extrapolation.

Our theoretical results have been derived under the assumption that UV
convergence has been reached. It would also be useful to know
finite-basis-size corrections in the opposite regime where IR
convergence is established, for instance through calculations in model
spaces with sufficiently small $\hw$. The remaining UV corrections
would, of course, depend on the interaction at hand. We have not yet
established a theoretical derivation but can resort to empirical
findings for SRG-evolved nucleon-nucleon interactions from chiral EFT
with an SRG parameter $\lambda$~\cite{bogner2008}.  We plotted IR
converged ground-state energies (computed at large values of $2\kinf
L$) for various light nuclei as a function of $\Lambda_{UV}$ and found
as in Refs.~\cite{haxton2000,coon2012} that the empirical formula
\beq 
\label{EUV}
E(\LamUV)=E_\infty +A_0 e^{-2\left({\Lambda_{UV}\over \lambda}\right)^2} 
\eeq 
works quite well~\cite{inprep}.  (In practice we allow $\lambda$ to be
a fit parameter to optimize the fit.)  This formula is consistent with
an empirically successful ansatz used for individual $\hw$ values
(e.g., see Ref.~\cite{bogner2008}).  However, Eq.~\eqref{EUV} allows
results with different $\hw$ to be fit all at once.

\begin{figure}[th-]
\includegraphics[width=0.4\textwidth]{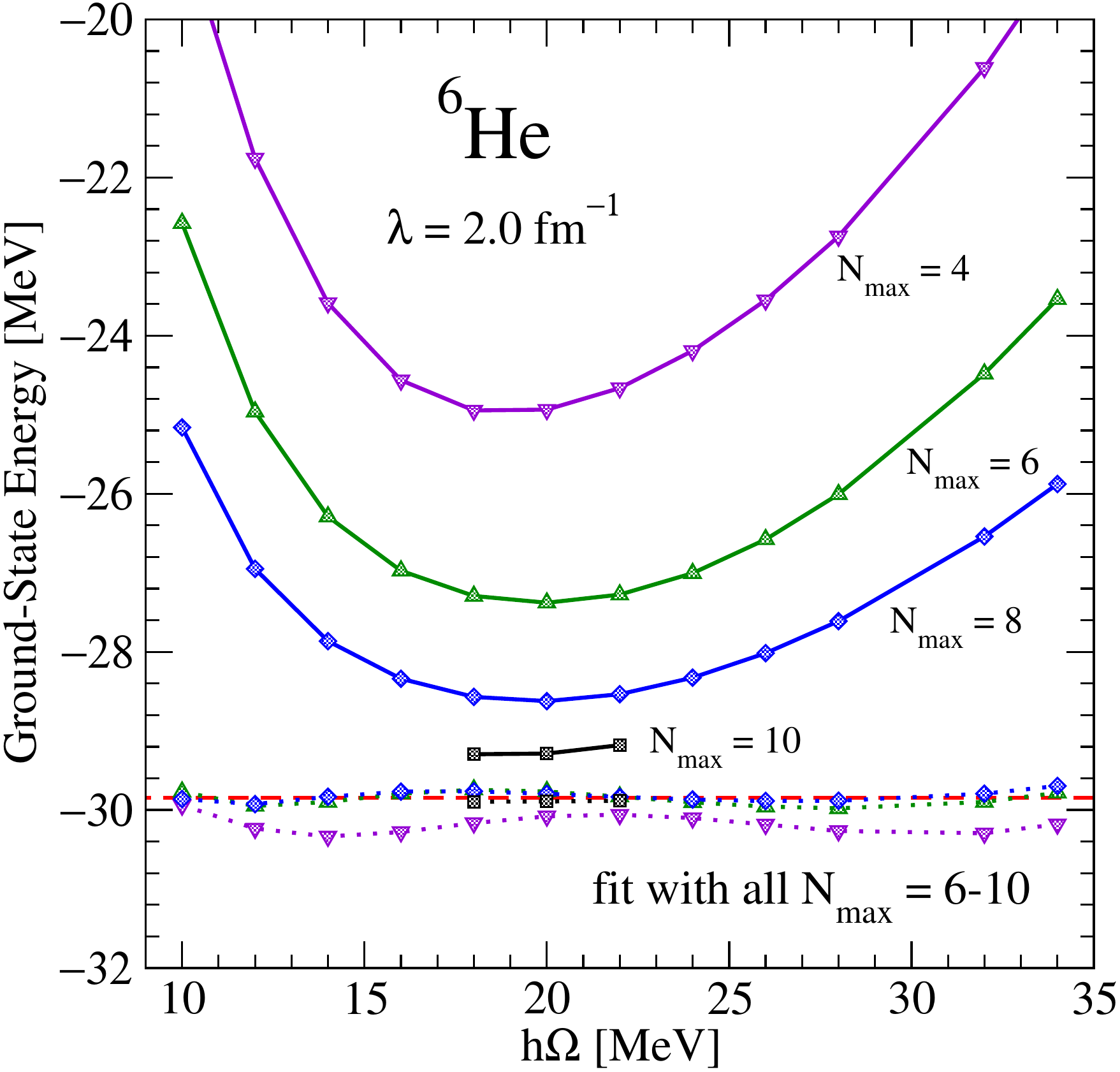}
\caption{(color online) Data points connected by full lines: NCFC
  ground-state energies of $^6$He. Red dashed line: Fit of
  Eq.~(\ref{Ecombi}) to NCFC data with $\Nmax=6$--10 yields $\Einf
  \approx -29.84\,$MeV. Data points connected with dotted line:
  Applying the correction of Eq.~(\ref{Ecombi}) to the NCFC data.
  ($N=\Nmax+1$)}
\label{fig6}
\end{figure}

If we combine the
empirical UV formula and the theoretically founded IR formula assuming
the corrections are approximately independent, then
\beq
\label{Ecombi}
 E(\LamUV,L)
   \approx \Einf + A_0 e^{-2\LamUV^2/A_1^2} + A_2 e^{-2\kinf L}
   \;.
\eeq
Here $\Einf$, $A_0$, $A_1$, $A_2$, and $\kinf$ are fit parameters that
are determined from a simultaneous optimization to data at all $\hw$,
including in the intermediate region where both IR and UV corrections
are significant.  The resulting value $\Einf \approx -29.84\,$MeV from using
all $\Nmax = 6$--10 points, which is in good agreement with the
IR-only fit in Fig.~\ref{fig4}, is plotted as a dashed red line in
Fig.~\ref{fig6}.  The points connected by dashed lines are obtained by
subtracting the corrections in Eq.~\eqref{Ecombi} from the NCFC
energies.  Thus a perfect fit would find all points lying on the line
for $\Einf$.  (Note: the $\Nmax$ values in the figure are for
\emph{excitations} above the ground state~\cite{bogner2008}, so
$N=\Nmax+1$ for $^6$He~\cite{coon2012,inprep}.)  All corrected points
included in the fit are close to the $\Einf$ line and even the
corrected $\Nmax=4$ energies (which were not included in the fit) 
are only slightly overbound.

In summary, we derived analytical results for the finite-basis-size
corrections of nuclear radii and energies that are valid in oscillator
spaces with converged ultraviolet physics. The computation of the
corrections is robust and appears to be applicable to halo nuclei.  In
combination with an empirical formula for the ultraviolet correction
for SRG-transformed interactions, consistent and much-improved
extrapolations of ground-state energies are possible.  The analytical
results can be extended to other long-range observables that are
sensitive to the tail of the nuclear density.  A systematic study of
the extrapolation procedure including an error analysis will be
presented in a future work~\cite{inprep}.


\begin{acknowledgments}
  We thank S. Bogner, S. Coon, U. Heinz, M. Kruse, P. Maris, W.
  Nazarewicz, R. Perry, and J. Vary for useful discussions.  This work
  was supported in part by the National Science Foundation under Grant
  No.~PHY--1002478 (The Ohio State University), and the Department of
  Energy under DE-FG02-96ER40963 (University of Tennessee) and
  DEAC05-00OR22725 (Oak Ridge National Laboratory). This research used
  resources of the Leadership Computing Facility at the
  Oak Ridge National Laboratory.
\end{acknowledgments}

\end{document}